\documentclass{aastex}
\bibpunct{[}{]}{;}{a}{}{,} % this generates square bracket references

\begin{document}

\title{A Radio Telescope Search for Axions}

\author{B. D. Blout\altaffilmark{1}, E. J. Daw\altaffilmark{1},
	M. P. Decowski\altaffilmark{1}, Paul T. P. Ho\altaffilmark{2},
	L. J Rosenberg\altaffilmark{1},
	and D. B. Yu\altaffilmark{1}}

\altaffiltext{1}{Department of Physics and Laboratory for Nuclear Science, 
	Massachusetts Institute of Technology, Cambridge, MA 02139}
\altaffiltext{2}{Center for Astrophysics, Harvard University,
	Cambridge, MA 02138}

\begin{abstract}
The axion is a hypothetical elementary particle and a cold dark matter 
candidate.  It could dominate potential wells of most astrophysical 
objects.  Axions spontaneously decay into nearly monochromatic microwave 
photons.  We present results from a radio telescope search for these 
axion decay photons of mass $m_a = 298$ to 363 $\mu\mathrm{eV}$ in Local 
Group dwarf galaxies.  We report a limit on the axion-to-two-photon 
coupling constant
$g_{a\gamma\gamma} > 1.0\times10^{-9}$ $\mathrm{GeV}^{-1}$.
\end{abstract}


\section{Introduction}
The axion is a well-motivated hypothetical elementary particle.  It is
the pseudo-Goldstone boson associated with a Peccei-Quinn symmetry
invoked to solve the strong-CP problem in QCD \cite{pq}
\cite{weinberg}.  The axion is a good cold dark matter candidate
\cite{k&t}, it could today dominate potential wells of most
astrophysical objects.  The axion spontaneously decays into two photons
with lifetime \cite{k&t}
\begin{equation}
\label{lifetime}
\tau_{a} = 6.8\times10^{24}
\left(\frac{0.72}{E/N-1.95}\right)^2\left(m_a/\mathrm{eV}\right)^{-5}\mathrm{s}\quad ,
\end{equation}
where $E/N$ is a model-dependent parameter.  The search described 
here seeks a spectral line  from axion decays in halos of 
Local Group dwarf galaxies.

There is a model range of perhaps a factor of 10 for the
axion-to-two-photon coupling constant $g_{a\gamma\gamma}$
\cite{kim:1}.  In two very different benchmark models,
$g_{a\gamma\gamma}$ varies from about
$1.35\times10^{-16}(m_a/\mu\mathrm{eV)GeV}^{-1}$ (the DFSZ model
\cite{zhit}) to $3.65\times10^{-16}(m_a/\mu\mathrm{eV)GeV}^{-1}$ (the
KSVZ model \cite{kim:2}).  Various terrestrial experiments and
astrophysical/cosmological arguments constrain the axion mass to the
approximate window $10^{-6}$ to $10^{-3}$ eV, and
$g_{a\gamma\gamma}$ to less than $\sim1\times10^{-10}$
$\mathrm{GeV}^{-1}$ \cite{turner}.  In one such astrophysical argument,
leading to the ``red giant limit''\cite{raffelt},
axions alter energy transport in
red giants.  An axion with $g_{a\gamma\gamma}$ greater than
$1\times10^{-10}$ $\mathrm{GeV}^{-1}$ would noticeably alter the
abundances in the Hertzsprung-Russell horizontal branch.
This limit, while the most sensitive in the upper two decades of the
allowed axion mass window, does not reach the benchmark model sensitivities.
Further, this limit depends on detailed models of red giant stellar
evolution.

By contrast, the main assumption for this radio telescope search is that
axions dominate potential wells.  This search examines Local
Group dwarf galaxies.  If axions are present in these galaxies, then
photons from their decay will appear as narrow lines in radio telescope
power spectra \cite{bershady}.

\section{Expected Signal}
\label{expectedSignal}
An estimate of the expected axion signal requires three main input
parameters: the spatial distribution of dark matter in a potential well,
the particular axion model, and the axion mass.  Each halo axion can
decay into two photons, each with frequency $\nu=(1/2)m_a
c^2/h$ in the axion rest frame.  The volume power density from axion
decay in the halo is $P_V=(1/2)\rho_a c^2/\tau_a$ where $\rho_a$ is the
axion mass density and $\tau_a$ is the axion lifetime.  Projecting
$P_V$ onto the sky gives the power per solid angle
$P_S(\Omega)$.

The observed power from axion decays is then given by convoluting $P_S(\Omega)$
with the beam pattern $B(\Omega)$
integrated
over the surface $A$ of the antenna
\begin{equation}
\label{powerFlux}
P_{obs}=\int{\frac{1}{4\pi D^2} P_S(\Omega) B(\Omega) dA d\Omega}\quad ,
\end{equation}
where D is the distance from Earth to the galaxy.  For the Haystack
antenna, $B(\Omega)$ is a two-dimensional Gaussian with beamwidth
$1.2\times\mathrm{7mm/36.6m}$.  The factors in equation
(\ref{powerFlux}) lead to the relation
$P_{obs}\propto m_a^{\alpha}$ with
$3<\alpha<5$ \cite{alpha}.  The value of the scaling parameter $\alpha$
depends on the source size and beamwidth.  An axion spectral line would
appear with a Doppler-broadened width $\Delta\nu/\nu\sim\sigma_{LOS}/c$
where $\sigma_{LOS}$ is the line-of-sight velocity dispersion within the
object.

We modeled the halo mass distribution as an isothermal sphere
\cite{chandra} with radial mass density
\begin{equation}
\label{massDistribution}
\rho(r)=\frac{\sigma_{LOS}^2}{4\pi G (r^2+a^2)}\quad ,
\end{equation}
where $G$ is the gravitational constant, $r$ is the radial distance from
the object center, and $a$ is the core radius \cite{binney&tremaine}.
We use a cutoff radius of $10a$ to avoid divergent total mass; these
results are only very weakly dependent on the choice of cutoff radius.
There are other models with a steeper mass falloff \cite{king}, the
isothermal sphere model leads to the more conservative limit on coupling
constant.

We chose three Local Group dwarf galaxies for study: Pegasus, Leo I, and
LGS 3 \cite{mateo}.  Dwarf galaxies were chosen because they have low virial
velocities, resulting in a narrow signal line,
$\delta\nu/\nu\sim10^{-5}$, easily distinguishable from broader
instrument-induced structure.  Also, these objects have high
mass-to-light ratios, and are likely dominated by dark matter.  Further,
these three galaxies are not known to be tidally disrupted \cite{burkert},
and the angular diameter of these
galaxies --- several acrminutes --- is well matched to the
antenna beamwidth.

\section{Observations}
Observations were made at Haystack Observatory (Westford, MA) on the
37-meter radio telescope \cite{haystack} at frequencies from 35.92 to
44.08~GHz.

Each observation is an ``off/on'' background-subtracted power spectrum
with a 160~MHz bandwidth.  Each off/on spectrum was derived from two
measurements.  First, the telescope was pointed a small angle ahead of
the object as it moved across the sky and a 30 second ``off source''
background spectrum recorded.  This off source region of the sky has no
known radio sources \cite{NED},
and represents the same section of the atmosphere and radome
which will shortly be tracked with the target source.
Second, the telescope was
pointed at the object and a 30 second ``on source'' spectrum recorded.
Finally, the off source spectrum subtracted from the on source spectrum
yielded the off/on spectrum.  This off/on technique, in tracing the same
arc of the sky and radome
twice, minimizes terrestrial artifacts.

Before making an observation, the antenna temperature was calibrated
with a blackbody load placed in front of the receiver feed horn.  This
calibration was repeated for every change in frequency, change in
observing object, and otherwise approximately every 10 minutes.
This serves to calibrate the instrumental response
as a function of frequency.
Typical system temperatures were in the range 100--230~K.

There are two major sources of error.
The first is ``pointing error'', the angular mismatch between
commanded and actual antenna direction.
From known sources, such as planets and quasars,
the estimated pointing error varied from
negligible to 15 arcsec.
For this analysis, we used a conservative fixed pointing
error of 22 arcsec.  A fully realized pointing error results in
a signal power degradation of 35\%,
our dominant uncertainty.

The second source of error is uncertainty in the
``efficiency'' of the telescope, which represents
all sources of signal attenuation.
These include the atmosphere, scattering from
the radome, and varying gravational deformation
of the antenna as the dish moves in elevation.
Again from known sources, the estimated efficiency
is between 29\%\ and 34\%.
For this analysis, we conservatively use a constant
efficiency of 27\%.

\section{Results}
Power spectra were collected at the Haystack Observatory on 21
April 99, 15 October 99, 29-31 December 99, and 21 March 00.  The total
number of spectra at any one frequency is at least three and sometimes
more than ten.  Higher numbers of observations were made for
frequencies that contained unusual features or candidates.  (As
explained below, these have all been ruled out as axions.)

Each spectrum has a 160~MHz bandwidth and was derived from a
256-time-lag autocorrelation spectrometer.
Thus, each bin has Nyquist width 625~kHz; the expected
axion signal has a width of about two bins.  The high and low frequency
ends of some spectra show bandpass filter skirts, so the first and last
quarter of the bins are removed from each spectrum.  The remaining
80~MHz of data per spectrum are used for all subsequent analysis.

The baseline of each spectrum is estimated from a least-squares fit to a
$4^\mathrm{th}$-order polynomial.  This estimated baseline is then subtracted
from the spectrum.  In Monte Carlo simulations, the order of the fit
polynomial improves the axion signal-to-noise ratio up to
$4^\mathrm{th}$-order, after which the sensitivity changes little.

A typical background-subtracted power spectrum is shown in figure 1.
The vertical axis shows power spectral density, the horizontal axis
shows frequency.  The ends of the spectrum have been removed, leaving
the central 128 of the original 256 bins.  The curve shows the fitted
baseline.  In addition, a simulated axion of coupling strength
$g_{a\gamma\gamma}=1.2\times10^{-9}$ $\mathrm{GeV}^{-1}$ has been
superimposed near the middle of the spectrum.

Bins from different spectra at the same frequency are combined
into a weighted average \cite{combining},  forming a ``combined
spectrum''.  The combined spectrum for dwarf galaxy LGS 3 is shown in
figure 2.  The spectrum shows the relatively
large system noise temperature near 39~GHz.

Next, the combined spectrum was normalized using its noise temperature
\cite{combining}.  Figure 3 is a histogram of the deviations in the LGS
3 normalized combined spectrum.  The line shows the expectation for unit
variance Gaussian noise.

We formed a ``candidate spectrum'' by averaging every two adjacent
bins of the normalized combined spectrum.  Candidates are defined as
those bins in the candidate spectrum exceeding a selected threshold.
The filled circles in figure 4 show the number of candidates (left
vertical axis) versus threshold for LGS 3.  The horizontal axis is
threshold in units of the normalized combined spectrum's standard
deviation.  The curve shows the expected number of candidates versus
threshold for unit variance Gaussian noise.

From figure 4
we selected a threshold of 2.3, resulting in 16 candidates.  In
Monte Carlo simulations we injected axion peaks with varying power and
the expected virial width into the unfitted spectra.  We determined that
an axion of power of about $10^{-18}$ W gave a 96\% search confidence at
the selected threshold; this is the search power sensitivity,
$P_{sens}$.  The injected power varied slightly across the spectrum to
maintain a flat 96\% search confidence.  The open circles in figure 4
show the search confidence for axions injected with $P_{sens}$.  Like
the red giant limit, this power sensitivity is much higher than that
required to detect KSVZ axions.

An axion signal should always be present at a given frequency.  We
returned to Haystack Observatory on 21 Mar 00 to examine the 16
candidates from LGS 3.  We determined that in this later data the
corresponding power for each candidate was small, this is consistent
with the 16 candidates being noise fluctuations.  We therefore
eliminated all 16 candidates while maintaining overall search confidence
at 96\%.

$P_{obs}$ for all three galaxies are approximetely
the same.  However, the
smaller angular size of LGS 3 yields better coupling constant
sensitivity \cite{alpha}.  The search procedure for Leo I and Pegasus is
identical to LGS 3, except the power threshold is set high enough to
exclude any candidates.

\section{Coupling constant sensitivity}
The relation between measured and predicted (KSVZ) power and coupling
constant sensitivity is
\begin{equation}
\frac{P_{sens}}{P_{KSVZ}}=\left(\frac{g_{sens}}{g_{KSVZ}}\right)^{\alpha}
\quad ,
\end{equation}
where $P_{sens}\mbox{ and }g_{sens}$ are the axion power and
coupling constant sensitivity of this search.  $\alpha$ is that of section
\ref{expectedSignal} and $P_{KSVZ}$ is obtained from equation
\ref{powerFlux}.  Figure 5 shows $\alpha$ as a function of frequency for
LGS 3, Leo I and Pegasus.
As frequency increases, the beamwidth of the antenna narrows,
the source looks relatively more diffuse, and $\alpha$ decreases.

The excluded values of the axion-to-two-photon coupling constant at 96\%
search confidence for LGS 3, Leo I, and Pegasus are shown in figure 6.
The lower horizontal axis is the photon frequency, and the upper
horizontal axis is the corresponding axion mass.  The vertical axis
shows the excluded coupling constant $g_{a\gamma\gamma}$.  The
sensitivity varies with the receiver noise and source.  We excluded
axion-to-two-photon coupling constants of
$1.0\times10^{-9}\mathrm{GeV}^{-1}$ or greater.

\section{Conclusions}
This search ruled out axions of mass 298 to 363 $\mu\mathrm{eV}$ with
axion-to-two-photon coupling of
$g_{a\gamma\gamma}>1.0\times10^{-9}$ $\mathrm{GeV}^{-1}$ at 96\%
confidence.  For the first time, a radio telescope was used to search
for cold dark matter axions in the allowed mass window.

The sensitivity of this result is comparable to the red giant limit.
However, the assumptions are very different; the radio telescope 
technique does not rely on detailed models of stellar evolution.

The RF-cavity axion search
technique \cite{sikivie} achieves remarkable sensitivity at the
lower axion masses.
Radio telescopes are more suited to search for axion
of higher mass.
Present radio telescope technology allows searches to the
very upper end of the allowed axion mass window
at sensitivities comparable to this result.

\acknowledgments
\begin{center}
{\bf Acknowledgements}\\
\end{center}
We are grateful to the research staff of Haystack Observatory for the
use of the facility and helpful discussions.  In particular, we thank 
J. Ball, P. Pratap and P. A. Shute for their guidance.
We also thank J. Hewitt for her many helpful suggestions.
Support for the use of the Haystack telescope was provided
by the National Science Foundation under grant DUE-9952246.


\begin{thebibliography}{}
\bibitem{alpha} [10] The axion-to-two-photon coupling constant is proportional
to the axion mass.  If the source is very small (e.g., a point source),
then the beam pattern weighting in equation (\ref{powerFlux}) is
constant.  The observed power would then scale as $m^5$.  If the source
is diffuse, then equation (\ref{powerFlux}) approaches the integral of a
two-dimensional gaussian distribution.  In this case the beam pattern is
convoluted with the nearly constant $P_S(\Omega)$ resulting in two
additional factors of the wavelength.
\end{thebibliography}
\end{document}